\documentclass[12pt]{article}
\usepackage[hypertex]{hyperref}
\usepackage[dvipdfmx]{graphicx} 
\usepackage{graphicx,amsmath,amssymb,amsfonts,cite,bm}

\def\a{\alpha} \def\b{\beta} \def\g{\gamma}  \def\d{\delta}        \def\l{\lambda}  \def\m{\mu} \def\n{\nu}     \def\r{\rho}       \def\o{\omega} 

\def\dg{\dagger} \def\del{\partial} \def\nn{\nonumber}

\newcommand{\lsp}{ \left ( }
\newcommand{\rsp}{ \right ) }
\newcommand{\Lg}{\mathcal{L}}
\newcommand{\we}{\wedge}

\newcommand{\getsto}{\leftrightarrow}

\newcommand{\vev}[1]{ \langle {#1} \rangle }

\newcommand{\tr}{{\rm tr}}
\newcommand{\Tr}{{\rm Tr}}

\newcommand{\diag}[2]{ \begin{pmatrix}  #1 & 0 \\ 0 & #2 \\   \end{pmatrix}  }
\newcommand{\offdiag}[2]{ \begin{pmatrix} 0 & #1 \\ #2 & 0 \\   \end{pmatrix} }

\begin{document}

\begin{titlepage}

\begin{flushright}
STUPP-17-230
\end{flushright}

\vskip 1.35cm

\begin{center}
{\LARGE \bf A possibility on prohibition of Higgs mass \\ by the extended Lorentz transformation \\ in noncommutative geometry}

\vskip 1.2cm

Masaki J. S. Yang

\vskip 0.4cm

{\it Department of Physics, Saitama University, \\
Shimo-okubo, Sakura-ku, Saitama, 338-8570, Japan\\
}

\begin{abstract} 

In this letter, we propose the extended Lorentz transformation 
in noncommutative geometry, 
as a possibility on prohibition of the Higgs mass. 

Since it is difficult to build the symmetry between the connections $A_{\m}$ and $H$, 
the transformation is defined for the differential two-forms. 
The parameter of the transformation $\o$ changes a two-form 
into other two-forms. 
Comparing the coefficients of the two-forms, 
the transformations are translated to those of the product fields 
$F_{\m\n}, D_{\m} H$ and $HH^{\dg}$. 
It shows the invariance of the bosonic Lagrangian explicitly. 

\end{abstract} 

\end{center}
\end{titlepage}

\section{Introduction}

The Higgs model in the noncommutative geometry (NCG), proposed by Connes and Lott \cite{Connes:1990qp}, is an interesting possibility of the explanation of the Higgs boson. 
In this picture, the Higgs boson is interpreted as a gauge boson along the discrete fifth dimension, 
which has the noncommutative differential algebra. 
This concept is applied in various theories, {\it e.g.,}  
grand unified theory (GUT) \cite{Chamseddine:1992kv,Chamseddine:1993is,Yang:2015gsa}, 
ideas related to the extra dimensions \cite{Alishahiha:2001nb}, supersymmetry \cite{Chamseddine:1994np}, and so on.

A crucial problem in this model is that 
the Higgs mass cannot be forbidden by the gauge invariance. 
As a straightforward idea, a deformed five dimensional Lorentz symmetry should play such a role.
In this letter, we propose the extended Lorentz transformation 
in noncommutative geometry, 
as a possibility on prohibition of the Higgs mass. 
Since it is difficult to build the symmetry between the connections $A_{\m}$ and $H$, 
the transformation is defined for the differential two-forms. 
The parameter of the transformation $\o$ changes a two-form 
into other two-forms. 
Comparing the coefficients of the two-forms, 
the transformations are translated to those of the product fields 
$F_{\m\n}, D_{\m} H$ and $HH^{\dg}$. 
It shows the invariance of the bosonic Lagrangian explicitly. 

For the Lagrangian of fermions, the appropriate matrix representation of this transformation is not be found 
\footnote{In the first version of ArXiv, we suggest the invariance of the Lagrangian of fermions by a gamma matrix-like operator  $\hat \g^{M}$ 
whose action is defined by the commutation relation, $\hat \g^{M} X = [\g^{M} , X] $. }.
The form of the transformation indicates that 
the representation space of fermions is twice larger than one of gauge connections. 
Such a representation might be realized by the real structure $\Psi = (\psi, \psi^{c})$ \cite{Connes:1995tu, Okumura:2001jz} or related concepts. 

Finally, we comment on the Coleman--Mandula theorem \cite{Coleman:1967ad}. 
The theorem is usually interpreted as prohibiting the symmetry between the Minkowski space and the internal space. 
However, the extended Lorentz transformation is defined in the five dimensional noncommutative space. It should be broken (for example, at the Planck scale) to the direct product of the Lorentz group and the gauge groups to produce the finite Higgs mass. 
Then, the symmetry  does not contradict to the theorem which is applied in the broken phase of the extended Lorentz symmetry. 
Similar discussions can be found in the graviweak theory \cite{Nesti:2007ka}.


\section{Higgs model in the noncommutative geometry}

To describe this Higgs model, there are several formalizations to represent the  noncommutative differential algebra.  
The original formalism \cite{Connes:1990qp} and its succeeding papers \cite{Lizzi:2000bc, Chamseddine:2006ep, Chamseddine:2013rta} 
 (and for reviews, Refs.~\cite{ORaifeartaigh:1998pk, Lizzi:2008dc, Martin:1996wh})
utilizes the ``universal differential algebra'', which is a generalization of usual differential forms. 
Meanwhile, some of other formalizations are based on a more simple algebra, 
such as $dy y = - y dy$ \cite{Coquereaux:1990ev, Sitarz:1993zf, Morita:1993xj, Okumura:1996ez, Okumura:1997qi, Konisi:1998ur}. 

Here we shortly review the theory with the latter algebra in the simplest  $M^{4} \times Z_{2}$ model.
The extended  exterior derivative is defined as 
\begin{align}
 \bm d f \equiv df + d_{5} f \equiv \del_{\m} f dx^{\m} + [M, f] dy^{5},  
 \label{extder}
\end{align}
where $M_{nm} (n,m = L,R)$ is the {\it distance matrix} which defines vacuum expectation value (vev) 
and the mass of the Higgs boson. 
Since $M_{nm}$ is arbitrary parameters, the model still works with $M_{nm} = 0$.
It leads to the Higgs boson without vev and mass \cite{Yang:2015zoa}. 
From now on, $M = 0$ and $\bm d = d$ is assumed. The nilpotency of $\bm d$ is manifest.

The wedge products of one-forms $dx^{\m}$ and $dy$ are given by \cite{Okumura:1996ez},
\begin{align}
dx^{\m} \we dx^{\n} = - dx^{\n} \we dx^{\m}, ~~~
dx^{\m} \we d y^{5} = - d y^{5} \we dx^{\m}, ~~~
dy^{5} \we dy^{5} \neq  0.  \label{2}
\end{align}
The generalized connection $\bm A (x)$ is defined to be
\begin{align}
\bm A (x) = 
\begin{pmatrix}
A_{L \m} (x) dx^{\m} & H^{5} (x) dy_{5}  \\
H^{\dg}_{5} (x) dy^{5} & A_{R \m} (x) dx^{\m}  \\
\end{pmatrix} 
\equiv 
\begin{pmatrix}
A_{L} & H \\ H^{\dg} & A_{R} 
\end{pmatrix} ,
\end{align}
where $dy_{5} = - dy^{5}$ and $H_{5} = - H^{5}$ with $g_{MN} = (+ , -,-,-,-).$
The gauge transformation of $\bm A$ is given by
\begin{align}
\bm A' = 
\diag{G_{L}}{G_{R}}
\begin{pmatrix}
A_{L} & H  \\
H^{\dg} & A_{R} \\
\end{pmatrix}
\diag{G_{L}^{\dg}}{G_{R}^{\dg}}
+ 
\diag{d G_{L} \cdot G_{L}^{\dg}}{ d G_{R} \cdot G_{R}^{\dg}} .
\label{gaugetrf}
\end{align}
where $G_{L,R} (x) = \exp[ i t^{a} \a_{L,R}^{a} (x) ]$ are the unitary matrices of gauge transformations.  
Note that the Higgs field $H^{5} (x)$ transform as a bifundamental field in this model. 
Henceforth, we omit the argument $x$ if there is no confusion. 

The extended field strength is defined as
\begin{align}
\bm F &= \bm d \bm A + \bm A \we \bm A = 
\begin{pmatrix}
F_{L} + H^{5} H^{\dg}_{5} \, dy_{5} \we dy^{5} & D_{\m} H^{5} \, dx^{\m} \we dy_{5} \\
D_{\m} H^{\dg}_{5} \, dx^{\m} \we dy^{5} & F_{R} + H^{\dg}_{5} H^{5} \, dy^{5} \we dy_{5}
\end{pmatrix} .
\label{fieldF}
\end{align}
Here, $F_{L,R} = d A_{L,R} + A_{L,R} \we A_{L,R}$ and 
$D_{\m} H^{5} = \del_{\m} H^{5} + A_{L} H^{5} - H^{5} A_{R}. $
In order to build the gauge-invariant Lagrangian, 
we use the following inner products of two-forms 
 \cite{Morita:1993dn, Morita:1993zv}
\begin{align}
\vev{dx^\mu \we dx^\nu, dx^\rho \we dx^\sigma} &=g^{\mu\rho}g^{\nu\sigma}- g^{\mu\sigma}g^{\nu\rho},  \label{hodge1} \\
\vev{dx^\mu \we dy^{5}, dx^\nu \we dy^{5}} &=-\a^2 g^{\mu\nu} , \\
\vev{dy^{5} \we dy^{5}, dy^{5} \we dy^{5}}&= 2  \b^{4} \label{hodge3} ,
\end{align}
while other products between the two-forms to be vanish.
Summarizing these results, the bosonic Lagrangian is found to be
\begin{align}
\Lg_{B} &= - {\rm Tr} \vev{\bm F , \bm F} \nn \\
 &= - {1\over 2} \tr[ F_{L \m\n} F_{L}^{\m\n} + F_{R \m\n} F_{R}^{\m\n}] +
 \tr [2  \a^{2} | D_{\m} H^{5} | ^{2} - 4 \b^{4} | H^{\dg}_{5} H^{5} |^{2} ] , 
\label{bosonicL}
\end{align}
where $F_{(L,R) \, \mu\nu} = \lsp \partial_{\mu} A_{(L,R) \,\nu} - \partial_{\nu} A_{(L,R) \, \mu} + [A_{(L,R) \, \mu}, A_{(L,R) \, \nu} ] \rsp$. Tr and tr denote the trace over the external linear space and internal gauge spaces, respectively.  The gauge coupling constants are introduced by rescaling of fields, {\it c.f.} $A_{L \m} \to i g_{L} A_{L \m } . $ 

A crucial problem in this model is that 
the Higgs mass cannot be forbidden by the gauge invariance:  
In other words, 
\begin{align}
\Lg_{B}'  = \Lg_{B} + m^{2} \tr [ H^{\dg}_{5} H^{5}] ,
\label{Higgsmass}
\end{align}
 is also invariant under the gauge transformation. 
It is desirable if some symmetry prohibits the Higgs mass.  
As a straightforward idea, 
 a deformed five dimensional Lorentz symmetry should play such a role 
  in noncommutative geometry.

\section{Extended Lorentz transformation}

\def\A{\bm A}  \def\F{\bm F}

In this letter, we propose the infinitesimal transformation of such a symmetry, 
 the {\it extended Lorentz transformation}. 
At first,  we attempted to consider the following transformation:
\begin{align}
\A ' &= 
\begin{pmatrix}
1 & \o \\ - \o^{\dg} & 1 
\end{pmatrix}
\begin{pmatrix}
A_{L} & H \\ H^{\dg} & A_{R}
\end{pmatrix}
\begin{pmatrix}
1 & - \o \\  \o^{\dg} & 1 
\end{pmatrix}  
\\ &= 
\begin{pmatrix}
A_{L}  & H  \\ 
H^{\dg}  & A_{R} 
\end{pmatrix}
+ 
 \begin{pmatrix}
\o H^{\dg}  + H \o^{\dg} &
- \o A_{R}  + A_{L} \o  \\
- \o^{\dg} A_{L } + A_{R} \o^{\dg}  & 
- \o^{\dg}  H  - H^{\dg} \o 
\end{pmatrix} ,
\end{align}
or, in components, 
\begin{align}
\bm A'_{M} = 
\begin{pmatrix}
A_{L \m}  & H^{5}  \\ 
H^{\dg}_{5}  & A_{R}^{\m} 
\end{pmatrix}
+ 
 \begin{pmatrix}
\o_{\m}^{5} H_{5}^{\dg}  + H^{5} \o^{\dg}_{\m 5} &
- \o_{\m}^{5} A_{R}^{\m}  + A_{L\m} \o^{\m 5}  \\
- \o^{\m \dg}_{5} A_{L \m} + A_{R}^{\m} \o_{\m 5}^{\dg}  & 
- \o^{\m \dg}_{5}  H^{5}  - H_{5}^{\dg} \o^{\m 5} 
\end{pmatrix} . 
\label{Atrf}
\end{align}
In order to match the gauge transformation of each matrix element, 
the parameter $\o^{\m}_{5}$ should be a matrix $\o^{\m}_{5} = \o^{b \, \m}_{5} s^{b}_{ij}.$ 
The matrix $\o^{\m}_{5 ij} $ is transformed as $\o'{}^{\m}_{5 ij} = (G_{L} \, \o^{\m}_{5} \, G_{R}^{\dg})_{ij}$ under the gauge transformation~(\ref{gaugetrf}).
However, Eq.~(\ref{Atrf}) does not seem to make the Lagrangian invariant,  
because the anti-symmetrization of Eq.~(\ref{Atrf}) $\F_{MN} = \bm \del_{M} \A_{N}' + \A_{M}'  \A_{N}' - ( M \getsto N)$ can not be written by $F_{\m\n}$ and $D_{\m} H^{5}$. 

Then, we try to consider a similar transformation for the 
extended field strength $\F$. 
Here, the shortened notation is used
\begin{align}
\F \equiv 
\begin{pmatrix}
F_{L} + H \we H ^{\dg} & D H \\ D H^{\dg} & {\bm -} H^{\dg} \we H + F_{R}
\end{pmatrix} ,
\label{change-} 
\end{align}
with change of the sign of $H^{\dg} \we H$ in the 22 element.
Although it seems {\it ad hoc}, the later formula shows that 
this correction is required from consistent transformation of the Higgs field.  
The transformation of $\F$ is defined to be
\begin{align}
\F' &= 
\begin{pmatrix}
1 & \o \\ - \o^{\dg}  & 1
\end{pmatrix}
\begin{pmatrix}
F_{L} + H \we H ^{\dg} & D H \\ D H^{\dg} & - H^{\dg} \we H + F_{R}
\end{pmatrix}
\begin{pmatrix}
1 & - \o \\ \o^{\dg}  & 1
\end{pmatrix} ,  \label{FTrf} \\ 
 \d \F  & = 
\begin{pmatrix}
\o D H^{\dg} + DH \o^{\dg} & \o ( - H^{\dg} \we H + F_{R} ) - (F_{L} + H \we H^{\dg} ) \o \\
- \o^{\dg} (F_{L} + H \we H^{\dg}) + (- H^{\dg} \we H + F_{R}) \o^{\dg} & - \o^{\dg} DH - DH^{\dg} \o
\end{pmatrix} .
\label{Ftrf}
\end{align}

If this kind of transformation exists, 
the invariance of the bosonic Lagrangian can be shown easily 
by the trace cyclicity: 
\begin{align} 
\d \Lg & = - \Tr \vev{\d \bm F , \bm F } - \Tr \vev{\bm F , \d \bm F } 
= - 2 \Tr \vev{\d \bm F , \bm F } ,  \\
\Tr \vev{\d \bm F , \bm F } 
&= \tr \vev{ ( \o D H^{\dg} + D H \o^{\dg}) , (F_{L} + H \we H^{\dg})}  \nn \\
&+ \tr \vev{ \o ( F_{R} - H^{\dg} \we H) - (F_{L} + H \we H ^{\dg}) \o, DH^{\dg}} \nn \\
& + \tr \vev{ - \o^{\dg} ( F_{L} + H \we H^{\dg}) + (F_{R} - H^{\dg} \we H) \o^{\dg}, DH} \nn \\
&+ \tr \vev{ ( - \o^{\dg} D H - D H^{\dg} \o) , (F_{R} - H^{\dg} \we H)} = 0 . 
\label{trcancel}
\end{align}

Since the transformation~(\ref{Ftrf}) is defined for the differential forms, 
the action of $\o$  expected to be 
\begin{align}
\o (dx^{\m} ) & = \o^{\m}_{5} dy^{5},  ~~~ \o(dy^{5}) = - \o^{\n}_{5} dx_{\n} .
\end{align}
However, when this relation is applied for the two-forms, 
anti-symmetric $dx^{\m} \we dy^{5}$ will mapped to 
non anti-symmetric $dy^{5} \we dy^{5}$. 
In order to correct this point, the action of $\o$ is defined 
\begin{align}
\o (A \we B) = (\o A) \we B + (-1)^{\del A} A \we (\o B) ,
\end{align}
where $\del A$ is $Z_{2}$ parity of form $A$ ($\del (dx^{\m}) = 1, \del (dy) = -1$).
The action of $\o$ is also symbolically represented by
\begin{align}
\o = \o_{\m 5} (dy^{5} \, dx^{\m * } - (-1)^{\del A} dx^{\m} \, dy^{5*}),
\end{align}
where $dx^{\m *}$ and $dy^{5 *}$ are the dual of each one-forms
$dx^{\m*} dx_{\n} = \d^{\m \n}, dy^{5*} dy_{5} = 1$.  
Specifically, the action for the two-forms are
calculated as
\begin{align}
\o C_{\m\n} dx^{\m}\we dx^{\n} &= 
\o^{\m}_{5} C_{\m\n} dy^{5} \we dx^{\n} + \o^{\n}_{5} C_{\m\n} dx^{\m}\we dy^{5} , \\
\o C_{\m5} dx^{\m}\we dy^{5} &= \o^{\m}_{5} C_{\m5} dy^{5} \we dy^{5} + 
\o_{\n 5} C_{\m 5} dx^{\m} \we dx^{\n} , \\
\o C_{55} dy^{5} \we dy^{5} &= \o_{\m}^{5} C_{55}  dx^{\m} \we dy^{5} - \o_{\n}^{5} C_{55}  dy^{5} \we dx^{\n} .
\end{align}
The action from righthand side is defined similarly:
\begin{align}
(A \we B) \o  = (A \o ) \we B + (-1)^{\del A} A \we (B \o ) .
\end{align}

By these definitions, components of Eq.~(\ref{Ftrf}) found to be
\begin{align}
 (D_{\m} H^{5} dx^{\m} \we dy_{5}) \o^{\dg} & = 
 D_{\m} H^{5} \o^{\m 5 \dg} dy_{5} \we dy_{5} - D_{\m} H^{5} \o_{\n 5}^{\dg} dx^{\m} \we dx^{\n} ,\\
\o^{\dg}  (D_{\m} H^{5} dx^{\m} \we dy_{5})  & =
 \o^{\m 5 \dg} D_{\m} H^{5}  dy_{5} \we dy_{5} - \o^{\dg}_{\n 5}  D_{\m} H^{5} dx^{\m} \we dx^{\n} ,\\
\o (D_{\m} H^{\dg}_{5} dx^{\m} \we dy^{5}) & = 
 \o_{\m 5} D^{\m} H_{5}^{\dg} dy^{5} \we dy^{5} - \o_{\n}^{5} D_{\m} H_{5}^{\dg} dx^{\m} \we dx^{\n} 
 , \\
(D_{\m} H^{\dg}_{5} dx^{\m} \we dy^{5})  \o  & = 
D_{\m} H_{5}^{\dg}  \o^{\m}_{5}  dy^{5} \we dy^{5} - D_{\m} H_{5}^{\dg} \o_{\n}^{5} dx^{\m} \we dx^{\n} , 
\end{align}
and 
\begin{align}
\o^{\dg} (H \we H^{\dg}) &= - 2 \o_{\m 5 }^{\dg} H^{5} H_{5}^{\dg} dx^{\m}\we dy^{5},  ~~~ 
(H \we H^{\dg}) \o  = - 2 H^{5} H_{5}^{\dg} \o_{\m5} dx^{\m} \we dy^{5} ,  \\
 \o (- H^{\dg} \we H ) &= - 2 \o_{\m 5} H_{5}^{\dg} H^{5} dx^{\m} \we dy^{5} , ~~~ 
 (- H^{\dg} \we H ) \o^{\dg}  = - 2  H_{5}^{\dg} H^{5} \o_{\m 5}^{\dg} dx^{\m} \we dy^{5} ,
\end{align}
and
\begin{align}
F_{L} \o &= 2 \, {1\over 2} F_{L \m\n} \o^{\n}_{5} dx^{\m} \we dy^{5},  ~~~ 
\o^{\dg} F_{L} = 2 \,  \o^{\n \dg}_{5} {1\over 2} F_{L \m\n} dx^{\m} \we dy^{5},  \\
\o F_{R} & = 2 \,  \o^{\n}_{5} {1\over 2} F_{R \m\n}  dx^{\m} \we dy^{5},  ~~~ 
F_{R} \o^{\dg} = 2 \, {1\over 2}  F_{R \m\n} \o^{\n}_{5} dx^{\m} \we dy^{5}.
\end{align}

Comparing the coefficients of the same two-forms, 
 the components in $\d \F$ are found to be
\begin{align}
\d ({1\over 2} F_{L \m\n}) &= - D_{\m} H^{5} \o_{\n 5}^{\dg} - \o_{\n}^{5} D_{\m} H_{5}^{\dg},  \label{trf1} ~~~~~
\d ({1\over 2} F_{R \m\n}) = + \o^{\dg}_{\n 5}  D_{\m} H^{5} + D_{\m} H_{5}^{\dg} \o_{\n}^{5}, \\
\d (H^{5} H^{\dg}_{5}) &= D^{\m} H^{5} \o_{\m 5}^{\dg} +  \o_{\m}^{5} D^{\m} H_{5}^{\dg}, ~~~~~
\d (- H^{\dg}_{5} H^{5}) = - \o_{\m 5}^{\dg} D^{\m} H^{5}  - D^{\m} H_{5}^{\dg}  \o_{\m}^{5} , \label{trf2}
\end{align}
and
\begin{align}
\d (D_{\m} H) 
& = 2 \lsp \o^{\n 5} {1\over2} F_{R \m\n} + \o_{\m 5} H_{5}^{\dg} H^{5} - {1\over2} F_{L \m\n} \o^{\n 5} +  H^{5} H_{5}^{\dg} \o_{\m}^{5} \rsp ,  \\
\d (D_{\m} H^{\dg}) 
& = 2 \lsp {1\over2} F_{R \m\n} \o^{\n \dg}_{5} + H_{5}^{\dg} H^{5} \o_{\m 5}^{\dg} -  {1\over2} \o^{\dg}_{\n 5} F_{L \m\n} +  \o_{\m}^{5 \dg} H^{5} H_{5}^{\dg} \rsp .  \label{trf6}
\end{align}
Substituting these relation into Eq.~(\ref{trcancel}), 
we can show the bosonic Lagrangian is invariant. 
The condition $\d \Lg = 0$ requires $\b^{4} = \a^{2} = 1$ in Eqs.~(\ref{hodge1}-\ref{hodge3}).
It is natural relation for the inner products with five dimensional Lorentz symmetry. 

Actually, the Lagrangian with $\b^{4} = \a^{2} = 1$
\begin{align}
\Lg_{B} &= - {\rm Tr} \vev{\bm F^{\dg} , \bm F} \nn \\
 &= - {1\over 2} \tr [  F_{L \m\n} F_{L}^{\m\n} + F_{R \m\n} F_{R}^{\m\n} ]+
 \tr [2 | D_{\m} H | ^{2} - 2 | H H^{\dg} |^{2} - 2 | H^{\dg} H |^{2} ] , 
\label{Lagprime}
\end{align}
is invariant under the transformation~(\ref{trf1}-\ref{trf6}) by the trace cyclicity:
\begin{align}
\d \Lg =& - \tr [\d F_{L\m\n} F_{L}^{\m\n}] - \tr [\d F_{R \m\n} F_{R}^{\m\n} ]
+ 2 \tr [ \d (D_{\m} H) D^{\m} H^{\dg}] + 2 \tr [D_{\m} H \d (D^{\m} H^{\dg})] \nn \\
& - 4 \tr[ \d (H H^{\dg}) H H^{\dg} ]- 4 \tr [\d (H^{\dg} H) H^{\dg} H ]\\
 =& + \tr [ 2 (  D_{\m} H^{5} \o_{\n 5}^{\dg} + \o_{\n}^{5} D_{\m} H_{5}^{\dg}) F_{L}^{\m\n}] 
- \tr [2 ( \o^{\dg}_{\n 5}  D_{\m} H^{5} + D_{\m} H_{5}^{\dg} \o_{\n}^{5}) F_{R}^{\m\n}] \nn \\
&+ 2 \tr  [( \o^{\n 5} F_{R \m\n} -  F_{L \m\n} \o^{\n 5} )  D^{\m} H^{\dg}]
+ 2 \tr [ D_{\m} H ( F_{R \m\n} \o^{\n \dg}_{5}  -  \o^{\dg}_{\n 5} F_{L \m\n} )] \nn \\
&+ 4 \tr  [(-  \o_{\m 5} H_{5}^{\dg} H^{5}  +  H^{5} H_{5}^{\dg} \o_{\m}^{5} )  D^{\m} H^{\dg}]
+ 4 \tr [D_{\m} H ( -  H_{5}^{\dg} H^{5} \o_{\m 5}^{\dg} +  \o_{\m}^{5 \dg} H^{5} H_{5}^{\dg} )] \nn \\
& - 4 \tr [(D^{\m} H^{5} \o_{\m 5}^{\dg} +  \o_{\m}^{5} D^{\m} H_{5}^{\dg}) H H ^{\dg} ]
+ 4 \tr [( \o_{\m 5}^{\dg} D^{\m} H^{5} + D^{\m} H_{5}^{\dg}  \o_{\m}^{5}) H ^{\dg} H ]
= \, 0.
\label{invariance}
\end{align}
Then, the bosonic Lagrangian~(\ref{bosonicL}) 
\begin{align}
 \Lg_{B} ' = - \Tr \vev{\bm F' , \bm F'} = \Lg_{B},  
\end{align}
is invariant under the transformation.

In the usual NCG theory, 
Higgs mass can not be prohibited by the symmetry of theory. 
This problem originates that 
the field strength~(\ref{fieldF}) with the Higgs mass term
\begin{align}
\tilde {\bm F} &= 
\begin{pmatrix}
F_{L} + H^{5} H^{\dg}_{5} \, dy_{5} \we dy^{5} & D_{\m} H^{5} \, dx^{\m} \we dy_{5} \\
D_{\m} H^{\dg}_{5} \, dx^{\m} \we dy^{5} & F_{R} - H^{\dg}_{5} H^{5} \, dy^{5} \we dy_{5}
\end{pmatrix} \\
& + 
\offdiag{m_{\m} H^{5} dx^{\m} \we dy_{5}}{m_{\m} H^{\dg}_{5} dx^{\m} \we dy^{5}}, 
\label{prob}
\end{align}
is also gauge invariant under Eq.~(\ref{gaugetrf}), 
with the constant vector $m_{\m}$. 
The {\it extended Lorentz transformation} Eq.~(\ref{FTrf}) or Eqs.~(\ref{trf1}-\ref{trf6}) (with gauge invariance) clearly prohibit the second term in Eq.~(\ref{prob}). 
Indeed, Eq.~(\ref{trf2}) shows the Higgs mass term will be transformed as 
\begin{align}
m^{2}  \tr [\d (H^{5} H ^{\dg}_{5})] = m^{2} \tr  [\d (H^{\dg}_{5} H^{5})] = 
m^{2} \tr [ D^{\m} H^{5} \o_{\m 5}^{\dg} +  \o_{\m}^{5} D^{\m} H_{5}^{\dg} ] ,
\end{align}
and it is not invariant. 
Note that the change of the sign in Eq.~(\ref{change-}) is required from the consistentcy of 
$\tr \, \d (H^{5} H_{5}^{\dg}) = \tr \, \d(H_{5}^{\dg} H^{5})$. 
We propose this transformation 
as a candidate of symmetry which prohibit the Higgs mass. 
In this case, the symmetry will be impose some relation between gauge coupling constants $g_{L,R}$ and Higgs self-coupling constants $\l$. 

\subsection{Discussions}

The transformations Eqs.~(\ref{trf1}-\ref{trf6}) are defined for the products fields. 
However, it is not clear whether Eqs.~(\ref{trf1}-\ref{trf6}) can be rewritten to the 
transformation of single fields $A_{\m}'$ and $H_{5}'$. 
The problem is that the uniqueness of the transformation is somehow lost. 
When we show the invariance in Eq.~(\ref{invariance}), 
the potential term of Higgs is rewritten as $2|HH^{\dg}|^{2} \to |HH^{\dg}|^{2} + |H^{\dg} H|^{2}$
in Eq.~(\ref{Lagprime}).
This modification is required by the invariance. However, 
the following inequality holds
\begin{align}
\tr [\d (H^{5} H_{5}^{\dg}) H^{5} H_{5}^{\dg} ]
&= \tr [(\o_{\r}^{5} D^{\r} H^{\dg}_{5} +  D^{\r} H^{5}  \o^{\dg}_{\r 5} )  H^{5} H_{5}^{\dg} ] \\
& \neq \tr [( \o^{\r \dg}_{5} D_{\r} H^{5} + D^{\r} H^{5 \dg} \o_{\r5})  H_{5}^{\dg} H^{5}] 
= \tr [\d(H_{5}^{\dg} H^{5}) H_{5}^{\dg} H^{5}] ,
\end{align}
and then the uniqueness of the transformation for $|H^{5} H_{5}^{\dg}|^{2}$ is lost. 

One solution of this point is redefining  
the transformation~(\ref{trf2}) to that of the product field $|H^{5} H_{5}^{\dg}|^{2}$:
\begin{align}
2 \d \, \tr [| H^{5} |]^{4} \equiv \tr [\d (H^{5} H_{5}^{\dg}) H^{5} H_{5}^{\dg}] + \tr [\d (H_{5}^{\dg} H^{5}) H_{5}^{\dg} H^{5}] ,
\end{align}
because the transformation is defined for product fields in the first place.

Finally, we comment on the Coleman--Mandula theorem \cite{Coleman:1967ad}. 
The theorem is usually interpreted as prohibiting the symmetry between the Minkowski space and the internal space. 
However, the extended Lorentz transformation is defined in the five dimensional noncommutative space. It should be broken (for example, at the Planck scale) to the direct product of the Lorentz group and the gauge groups to produce the finite Higgs mass. 
Then, this symmetry does not contradict to the theorem which is applied in the broken phase of the extended Lorentz symmetry. 
Similar discussions can be found in the graviweak theory \cite{Nesti:2007ka}.

\section{Conclusions and Discussions}

In this letter, we proposed the extended Lorentz transformation 
in noncommutative geometry, 
as a possibility on prohibition of the Higgs mass. 
Since it is difficult to build the symmetry between the connections $A_{\m}$ and $H$, 
the transformation is defined for the differential two-forms. 
The parameter of the transformation $\o$ changes a two-form 
into other two-forms. 
Comparing the coefficients of the two-forms, 
the transformations are translated to those of the product fields 
$F_{\m\n}, D_{\m} H$ and $HH^{\dg}$. 
It shows the invariance of the bosonic Lagrangian explicitly. 

For the Lagrangian of fermions, the appropriate matrix representation of this transformation is not be found.
The form of the transformation indicates that 
the representation space of fermions is twice larger than one of gauge connections. 
Such a representation might be realized by the real structure $\Psi = (\psi, \psi^{c})$ \cite{Connes:1995tu, Okumura:2001jz} or related concepts. 

{\bf Acknowledgements:} This study is financially supported by the Iwanami Fujukai Foundation, and
Seiwa Memorial Foundation. 


\end{document}